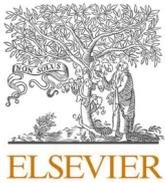
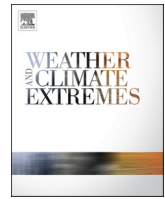

# Abrupt and persistent atmospheric circulation changes in the North Atlantic under La Niña conditions

Marina García-Burgos [a,*], Iñigo Gómara [b,a], Belén Rodríguez-Fonseca [a,c], Juan Jesús González-Alemán [a], Pablo Zurita-Gotor [a,c], Blanca Ayarzagüena [a]

[a] *Departamento de Física de la Tierra y Astrofísica, Universidad Complutense de Madrid, 28040, Madrid, Spain*
[b] *Departamento de Matemática Aplicada, Universidad de Valladolid, 40005, Segovia, Spain*
[c] *Instituto de Geociencias (IGEO), UCM-CSIC, 28040, Madrid, Spain*

## ABSTRACT

Several recent studies have linked the exceptional North Atlantic and Eurasian atmospheric evolution during late February and March 2018 to the Sudden Stratospheric Warming (SSW) that took place a few weeks earlier. February 2018 was characterized by an abrupt transition from the positive to the negative phase of the North Atlantic Oscillation (NAO) and a subsequent persistence of the negative NAO for several weeks. This paper investigates the contribution of atmospheric and oceanic phenomena to both the 2018 event and a set of 19 identified analogues (including the former) for the period 1959–2022. Evidence is given that La Niña conditions in the tropical Pacific and upstream North Atlantic cyclones play an important role as a trigger for these events. Ensuing two-way tropospheric-stratospheric coupling and eddy feedbacks provide extended-range persistence for negative NAO conditions. These results may help improve the prediction of such exceptional events.

**Key points**

- A Dynamical analysis of the exceptional North Atlantic-Eurasian atmospheric evolution of February 2018 and 18 identified precedents is performed.
- Oceanic and tropospheric processes are essential for the occurrence of abrupt and persistent atmospheric circulation changes in the North Atlantic.
- Eddy-feedbacks and troposphere-stratosphere coupling are found to be relevant for the negative NAO phase persistence.

## 1. Introduction

The large-scale atmospheric circulation variability in the North Atlantic is dominated by the North Atlantic Oscillation (NAO) at multiple timescales (Hurrell et al., 2003). Under a positive NAO phase, the polar jet is strengthened and shifted northward (Woollings et al., 2010), which is characterized by analogous changes in the North Atlantic stormtrack and a reinforcement of the subtropical high (Pinto et al., 2009; Gómara et al., 2014a). The opposite is observed under negative NAO (NAO-).

Evidence has been given that single extratropical cyclones can dramatically alter the circulation regime in the North Atlantic (Franzke et al., 2004; Rivière and Orlanski, 2007). Strong cyclones inducing large-scale Rossby wave breaking (RWB) events are typically involved in this process (Colucci, 1985; Lupo and Smith, 1995; Gómara et al., 2014b). Upstream cyclones in the North Atlantic preceding Scandinavian Blocking (SB), Greenland Blocking (GB) or NAO- weather regimes are clear examples of the former, where strong poleward advection of sub-tropical warm and moist air completely disrupts the dominant westerly flow in the North Atlantic (Michel and Rivière, 2011; Michel et al., 2012; Maddison et al., 2019). Conversely, eddy-forcing from ensuing upstream cyclones can contribute to the maintenance of the anomalous circulation (Nakamura and Wallace, 1993; Barnes and Hartmann, 2012). However, determining in which case the eddies force a new regime or maintain the existing one is not always simple.

Large-scale RWB and blocking events (e.g., SB) have been related to enhanced upward propagation of planetary waves into the stratosphere, leading to the weakening or even breakdown of the stratospheric polar vortex (Sudden stratospheric warmings - SSWs; Martius et al., 2009; Lee et al., 2019). During the ensuing 1–2 months after an SSW event, the downward propagation of the stratospheric anomalies can alter the North Atlantic tropospheric circulation, typically projecting on negative phases of the Northern Annular Mode (NAM) and the NAO (Baldwin and Dunkerton, 2001; Domeisen, 2019).

Ocean-atmosphere interactions can also play a role in the occurrence of persistent SB/NAO- states. The links between El Niño-Southern Oscillation (ENSO) and the NAO have also been analyzed (Fraedrich 1994; Gouirand and Moron 2003; Brönnimann 2007; Li and Lau 2012a, 2012b; Drouard et al., 2013, 2015) and appear to depend on ENSO flavor, seasonality, time period considered and the strength of the anomalies, which can also lead to non-linearities in these links (Fraedrich and Müller, 1992; Moron and Gouirand 2003; López-Parages and Rodríguez-Fonseca 2012; Frauen et al., 2014; Hardiman et al., 2019; Zhang et al., 2019; Trascasa-Castro et al., 2019; Weinberger et al., 2019; Jiménez-Esteve and Domeisen, 2020; Casselman et al., 2021). During late boreal winter, the extratropical atmospheric response to an El Niño event resembles the NAO- at surface levels (and vice-versa for La Niña; García-Serrano et al., 2011; Li and Lau, 2012b). Proposed ENSO-NAO teleconnection mechanisms encompass tropical, extratropical and stratospheric pathways (Brönnimann, 2007; Domeisen et al., 2019).

During late February 2018, an outstandingly swift and subsequently persistent weather regime transition took place in the North Atlantic (NAO+/SB/NAO- sequence - Vautard, 1990; González-Alemán et al., 2022; hereafter F18 event). Before the F18 event, persistent NAO+/SB conditions dominated the winter season. After F18, the atmospheric conditions in Europe dramatically changed. As a consequence, a severe cold spell causing exceptional snowfall with strong societal impacts affected large portions of Europe (Copernicus, 2018; Aon Benfield, 2019). In the following weeks, persistent rain and flooding conditions took place over Iberia and a dramatic decrease of precipitation affected northern Europe (Ayarzagüena et al., 2018; Drouard et al., 2019). In particular, increased rainfall over southwestern and central Europe ended the long drought that had affected the region for the previous two years (García-Herrera et al., 2019). The analysis of F18 poses great challenges, as many factors may have contributed to this transition. Firstly, La Niña Sea Surface Temperature (SST) conditions dominated the tropical Pacific since the previous autumn-winter. Further, two weeks before the F18 event, a major SSW took place (Ayarzagüena et al., 2018; Kautz et al., 2019).

In this study, the physical atmospheric and oceanic processes that potentially contributed to the occurrence of F18 are analyzed. The analysis is extended to 19 identified events (including F18) with analogous North Atlantic tropospheric characteristics to F18 occurring under La Niña conditions for the period October–March 1959–2022. Evidence is given in this article that physical processes linked to abrupt NAO transition events, such as air-sea interactions (Wills et al., 2016), cyclone development (Gómara et al., 2016), RWB (Benedict et al., 2004), etc., which have received relatively little attention so far in relation to the F18 event, were determinant for its triggering and that of its identified analogues. Overall, the stratosphere contributes to NAO-persistence, even though SSWs are only observed for a few events (Baldwin et al., 2021 and references therein).

## 2. Materials and methods

### 2.1. Data

The European Centre for Medium-Range Weather Forecasts Reanalysis v5 data (ERA5; Hersbach et al., 2020) were analyzed for the period October–March 1959–2022. ERA5 has 60 vertical levels with 0.25° longitude-latitude horizontal resolution, available every 6 h. The main variables utilized are sea level pressure (SLP), geopotential height ($z$), and zonal ($u$) and meridional ($v$) winds. The 1° longitude-latitude resolution SST data from the Met Office Hadley Centre (HadISST; Rayner et al., 2003) were also considered for the same period.

### 2.2. Extratropical cyclone considerations

An automatic tracking algorithm was utilized to identify extratropical cyclones (Murray and Simmonds, 1991; Pinto et al., 2005). The method considers the Laplacian of SLP as an indicator of a cyclone's geostrophic relative vorticity and provides full cyclone information (e.g., location, intensification rate). The method compares well with similar tracking schemes (Neu et al., 2013). Physically coherent cyclones were selected based on Pinto et al. (2009) criteria (cf. their *Methods* section). Additional constraints were applied to retain systems with specific tracks and intensities. The Normalised Deepening Rate (NDR; Sanders and Gyakum, 1980) was used to classify cyclones in terms of intensity: all cyclones (NDR >0 bergeron) and explosive cyclones (NDR ≥1 bergeron). NDR is defined as:

$$NDR = \Delta P/24 \cdot \sin 60 / \sin \varphi \qquad (1)$$

where $\Delta P$ is the pressure drop (hPa) and $\varphi$ the mean latitude (deg.) of the cyclone's surface center over a period of 24 h.

### 2.3. Rossby wave-breaking considerations

A daily two-dimensional RWB index was computed from ERA5 potential temperature ($\theta$) on the dynamical tropopause (2 Potential Vorticity Units surface). The index is the same as in Masato et al. (2012) and Gómara et al. (2014b). It provides two-fold information: local-instantaneous RWB occurrence (B index) and direction of breaking (cyclonic or anticyclonic; DB index). The index imposes a strict criterion on the latitudinal extension of the $\theta$ reversal (ca. 30° lat.) to identify large-scale local-instantaneous RWB events.

### 2.4. Climate indices and wave propagation

The daily NAO index was constructed based on monthly leading EOFs (Empirical Orthogonal Functions) of 500 hPa geopotential height (z500) anomalies [25°-80°N; 80°W-40°E] over the study period. This index was computed by projecting the corresponding z500 daily anomalies in the Euro-Atlantic sector onto the EOF1 pattern (Gómara et al., 2014a). A daily NAM index (Thompson and Hegerl, 2000) was calculated independently at each pressure level, following Baldwin and Dunkerton (2001), by projecting the geopotential height anomaly from a smoothed daily climatology on the leading EOF of deseasonalized, low-pass filtered (90-day moving average) geopotential field north of 20° N during the extended winter. Stratospheric propagation of planetary waves was assessed using the vertical component of the Eliassen-Palm flux (Edmon et al., 1980) for zonal waves k = 1–3 in the zonal mean. Similarly, the Plumb flux (Plumb, 1985) was computed for evaluating the propagation of the slowly-moving anomalies (5-day moving average) at a certain pressure level. The three-dimensional flux is given by the expression:

$$F_s = \frac{p}{10^5} \cos \varphi \begin{pmatrix} v^{*2} - \dfrac{1}{2\Omega \, a \sin 2\varphi} \dfrac{\partial (v^{*2} \varphi^*)}{\partial \lambda} \\ -u^* v^* + \dfrac{1}{2\Omega \, a \sin 2\varphi} \dfrac{\partial (v^{*2} \varphi^*)}{\partial \lambda} \\ \dfrac{f}{\left(\dfrac{\partial \check{T}}{\partial p} - \dfrac{\kappa \check{T}}{p}\right)} \left( v^* T^* - \dfrac{1}{2\Omega \, a \sin 2\varphi} \dfrac{\partial}{\partial \lambda} (T^* \varphi^*) \right) \end{pmatrix} \qquad (2)$$

where ∗ indicates the deviation from the zonal mean, $\lambda$ is the longitude, $\Omega$ is the Earth's rotation rate, $a$ is the Earth's radius, $\varphi$ is the geopotential, $\kappa$ is the Poisson constant, $p$ is the level pressure, $T$ is the temperature and $\check{T}$ indicates the average over the area north of 20°N of the temperature.

### 2.5. Calculation of anomalies, composites, and hypothesis testing

Daily anomaly fields were calculated by removing the corresponding daily long-term 1959–2022 mean in each grid-point. For the analysis of F18 analogues (section 3.2), composites of anomalies were computed.

Regarding the statistical significance of results, a Monte Carlo test with 1000 random permutations was applied. More specifically, composite maps of different atmospheric/oceanic variables such as SLP or wind at 200 hPa were computed for random samples that were picked with replacement from the whole dataset we were working with, i.e. the extended winter for the period 1959–2022. The values of all randomly generated composite maps in each grid point were used to generate a pdf of the statistics in that grid point. Once the pdf was generated, we determined the location of the actual mean value in the pdf. A confidence level of 95% (p-value<0.05) was considered throughout the whole analysis (95%).

## 3. Results

### 3.1. Analysis of the February 2018 event

The previous month to the late (22–29) February NAO abrupt change, the large-scale atmospheric-oceanic conditions over the North Atlantic were characterized by NAO + predominance, with an anomalous intense and persistent Azores anticyclone, warmer than usual SSTs in the latitudinal band from 20 to 40° N in the North Atlantic and La Niña conditions in the tropical Pacific (Fig. 1a).

The following 30-days' period was diametrically opposed in the North Atlantic, with negative SLP anomalies present over western Europe and positive over Scandinavia (NAO-), together with a weakening of the aforementioned North Atlantic and tropical Pacific SST anomalies (Fig. 1b). Iberia was affected by abundant precipitation in this period (Fig. 1b, green shading over land; Ayarzagüena et al., 2018; Drouard et al., 2019).

The apparent underlying mechanisms responsible for such dramatic circulation change took place the second half of February 2018, as made evident by the NAO index evolution in Fig. 1d, which dropped ∼5 standard deviation units (SDU; from NAO + to NAO-) in a couple of weeks. During these days, in particular on 19 and 24 February, a couple of North Atlantic upstream extratropical cyclones explosively intensified and triggered consecutive large-scale cyclonic RWB south of Greenland (Fig. 1c and S1; Kautz et al., 2019). These cyclonic RWB events promoted and subsequently reinforced advection of warm/moist subtropical air towards Scandinavia, causing the disruption of the westerly flow and the onset of SB (González-Alemán et al., 2022).

In addition to that, a SSW took place on February 12, 2018 (Ayarzagüena et al., 2018; Rao et al., 2018). This fact attests to the complexity of the F18 event, as multiple dynamical processes took place simultaneously (SSW, cyclone development, RWB, La Niña conditions, etc.). In a recent modeling study by González-Alemán et al. (2022), the key contribution of North Atlantic cyclogenesis in the F18 event was confirmed. To assess the generality of these results and the relative importance of the processes involved, events similar to F18 were subsequently identified from reanalysis data.

### 3.2. Identification of F18 analogues

The following criteria were applied to ERA5 data (October–March 1959–2022) to identify events with the most similar oceanic and tropospheric characteristics to F18.

(i) **Strong cyclonic RWB occurrence over Greenland:** More than 50% of points in the area [60°-75° N, 50°-20° W] (cf. Fig. 1c, solid rectangle) must return cyclonic RWB occurrence in a given day. The area corresponds to the region where the cyclonic RWB took place in F18.

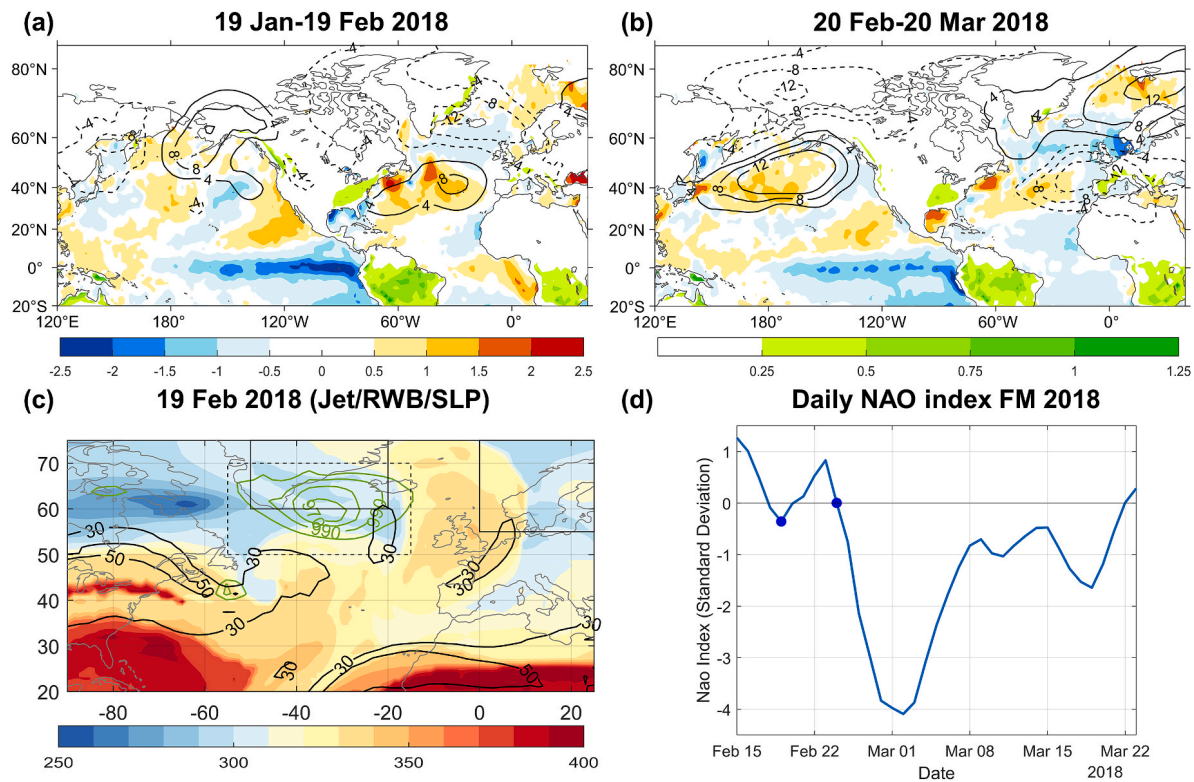

**Fig. 1. F18 description.** Seasonal SST (red/blue shadings over ocean areas, K), SLP (contours, hPa) anomalies and total precipitation (green shadings over land areas, above $10^{-2}$ m) for (a) 19 Jan-19 Feb 2018 and (b) 20 Feb-20 Mar 2018. (c) Potential temperature on the tropopause (2PVU, blue/red shadings in K), wind speed at 250 hPa (black contours in m s$^{-1}$) and SLP (green contours in hPa) on 19 February 2018. Boxes for event selection in Section 3.2 appear overlaid: cyclone occurrence [55°-15° W, 50°-70° N], cyclonic [50°-20° W, 60°-75° N] and anticyclonic [0°-25° E, 55°-75° N] RWB detection. (d) NAO index from 15 February-23 March 2018 (gpm SD$^{-1}$). Dots denote cyclone occurrences of 19 and 24 February. (For interpretation of the references to color in this figure legend, the reader is referred to the Web version of this article.)

(ii) **Abrupt NAO drop:** Considering lag 0 the first day the Strong cyclonic RWB (identified in (i)) takes place, the NAO daily index must drop more than 1 SDU between daily lags −2 to +10. The minimum NAO value during this period must be negative.

(iii) **NAO- persistence:** The daily NAO index must remain negative at least during 21 consecutive days after the abrupt NAO drop. The persistence starts to count the day the NAO index drops at least 1 SDU and turns negative. The criterion is based on the actual length of F18 event (26 days) and the characteristic timescale of individual eddy-forcing events (Rivière and Orlanski, 2007; Robert et al., 2017).

(iv) **La Niña:** By analogy with F18, only events with La Niña conditions in the Pacific (area averaged [5°N-5°S; 170°-120°W] SST anomalies lower than −0.5 K in the month of occurrence) were selected. Additional analyses (not included) show that the majority of events (19/35) occur during this ENSO phase (considering Niña, Niño and Neutral states) and that those occurring during non La-Niña phases exhibit different Northern Hemisphere large-scale circulation characteristics.

All these four criteria were imposed to precisely identify similar atmospheric/oceanic conditions to those of F18. However, conclusions do not significantly change if we slightly modify threshold values in criteria (ii) to (iii). Although some of these criteria might not be independent from the others (e.g., RWB occurrence and NAO state), they are necessary to obtain a very similar picture to F18 (motivation of the study).

### 3.3. Analysis of F18 analogues

Two subsets of analogues were constructed: one taking into account the NAO- persistence criterion (hereafter Persistent - P; 19 events) and the other without it, i.e. considering those events with less than 21 days of persistence (hereafter Not Persistent – NP; 77 events). P events persistence ranges from 21 to 69 days and NP from 1 to 20 days (Fig. S2a). The aim of generating two populations is to determine, on one hand, whether the mechanisms triggering the abrupt transition are the same regardless of the subsequent NAO- persistence and, on the other hand, the possible drivers of the latter. Relevant statistics for the two populations are provided in Table 1. Results show how explosive cyclone activity (NDR >1) is enhanced over the area [55°–15° W, 50°–70° N; dashed box Fig. 1c] the days prior and after strong cyclonic RWB is detected south of Greenland (i) for both populations. However, the total number of cyclones (NDR >0) around the RWB is only enhanced in P events. Subsequent anticyclonic RWB occurrence over Scandinavia appears significantly enhanced at daily lags −2 to +10 only for P events, consistent with the SB onset (Maddison et al., 2019). All these features corroborate the similarity of the P and NP events selected to F18 in terms of NAO behavior, cyclones involved and RWB activity in the North Atlantic. The difference between the populations relies on the following anticyclonic RWB over Scandinavia, increased for P events but with no significant changes for NP events. Nevertheless, there is no monthly preference with any of these samples nor a clear difference in their cyclones trajectories (Figs. S2b and S2c).

Fig. 2a depicts the environmental conditions present in the Northern Hemisphere before the abrupt change in NAO sign for P events (daily lags −7/-1 for z200 and SLP and lags −2/-1 for 250 hPa wind speed; the wind lag is shorter because the jet only accelerates a few days before the large-scale RWB). Two main circulation structures are observed. The first is evident in the North Atlantic. Positive SLP anomalies are located over the subtropical North Atlantic, together with negative anomalies between the Hudson Bay and Iceland. These conditions induce an accelerated SW-NE tilted jet stream near Newfoundland. This feature is a well-known precursor of strong upstream North Atlantic cyclones and subsequent SB/NAO- onset (Michel et al., 2012; Gómara et al., 2014b). Both SLP centers at the surface are the signature of two z200 anomalies that belong to a global wave pattern with wavenumber 5 (WN5) located at 45°N. This wave emerges from different regions of the Pacific and travels toward the North Atlantic (Fig. S3a). The second structure is a strong blocking event over the North Pacific, denoted by the strong positive SLP and z200 anomalies in that area (Fig. 2a). The positive z200 anomaly over the North Pacific (60°N) is part of a circumglobal wave with wavenumber 2 (WN2). These anomalies are known to enhance upward WN2-wave propagation into the stratosphere, due to constructive interference with climatological waves (Nishii et al., 2011). As a result, the stratospheric vortex can weaken and even lead to the occurrence of a SSW. Blocking events over the Northeastern Pacific have indeed been identified as preferred SSW precursors during La Niña winters (Barriopedro and Calvo, 2014). The described two main circulation structures observed for the P events, i.e. the North Atlantic anticyclone that speeds the jet and the WN2 at 60°N, are also clearly observed on the days preceding the F18 event (Fig. 1a).

A picture consistent with these results is that the blocking in the North Pacific and the anticyclonic RWB in the North Atlantic perturb the stratospheric polar vortex, by modifying the upward-propagating wave activity (Martius et al., 2009; Lee et al., 2019). The Plumb flux (Fig. 2c) confirms that the anomalous upward wave propagation originates in both the North Pacific and the North Atlantic regions. This suggests that the combination of both processes, namely the blocking in the North Pacific and the North Atlantic wave breaking, can significantly perturb the stratosphere and contribute to extended-range NAM-/NAO- persistence in the P subset. SSW occurrence, as determined by the Charlton and Polvani (2007) criterion, is significantly enhanced the weeks after P events (Table 1), but does not occur after every event (5/19 cases, Table S1).

Similar analyses were repeated for the NP subset (Fig. 3). Consistent with NP results in Table S1, the jet stream is again accelerated in the western North Atlantic from daily lags −2 to −1 (Fig. 3a), but it is more zonally oriented than in P (Fig. 2a). Overall, the blocking pattern is still observable, but its strength is significantly weaker compared to P (Fig. 2a and 3a). Additionally, the blocking now projects into a WN1 circumglobal wave at 200 hPa, instead of WN2.

The stratospheric contribution to the NAO+/NAO- transition also exhibits significant differences compared to P (Fig. 3b and 2b). There is some anomalous upward wave activity flux from the surface a few days before the NP dates, but its influence does not reach the stratosphere, nor it perturbs the vortex. The transition towards NAO- only appears to take place within the troposphere, although it is not fully captured by the NAM index which presents anomalies close to zero (white colors in Fig. 3b). The Plumb Flux map shows a significant upward wave center over the North Pacific, but again much weaker compared to P, and there is no appreciable vertical propagation south of Greenland (Fig. 3c). This may explain the unperturbed stratospheric vortex state for NP events.

**Table 1**
**Detailed information on events.** Frequencies of occurrence of atmospheric phenomena for specific regions and daily lags for climatology and identified events dates (5th and 6th columns). For each phenomena the same calculations as for the events were replicated considering 1 000 randomly selected dates from the climatological period (October–March 1959–2022). Results with 95% conf. int. (Monte Carlo test) in bold.

| Indicator | Lags | Region | Clim. | Persistent (19) | Non-persistent (77) |
|---|---|---|---|---|---|
| All cyclones south of Greenland (cyclones day$^{-1}$) | −7/+1 | [55°-15°W, 50°-70°N] | 0.7025 | **0.7953** | 0.6888 |
| Explosive cyclones south of Greenland (cyclones day$^{-1}$) | −7/+1 | [55°-15°W, 50°-70°N] | 0.1679 | **0.2456** | **0.1977** |
| Anticyclonic RWB over Scandinavia (RWB day$^{-1}$) | −2/+10 | [0°-25°E, 55°-75°N] | 0.0513 | **0.1404** | 0.0682 |
| Preceding major SSW (SSW day$^{-1}$) | −20/0 | Northern Hemisphere | 0.0035 | 0.0053 | 0.0039 |
| Posterior major SSW (SSW day$^{-1}$) | +1/+21 | Northern Hemisphere | 0.0035 | **0.0132** | 0.0052 |

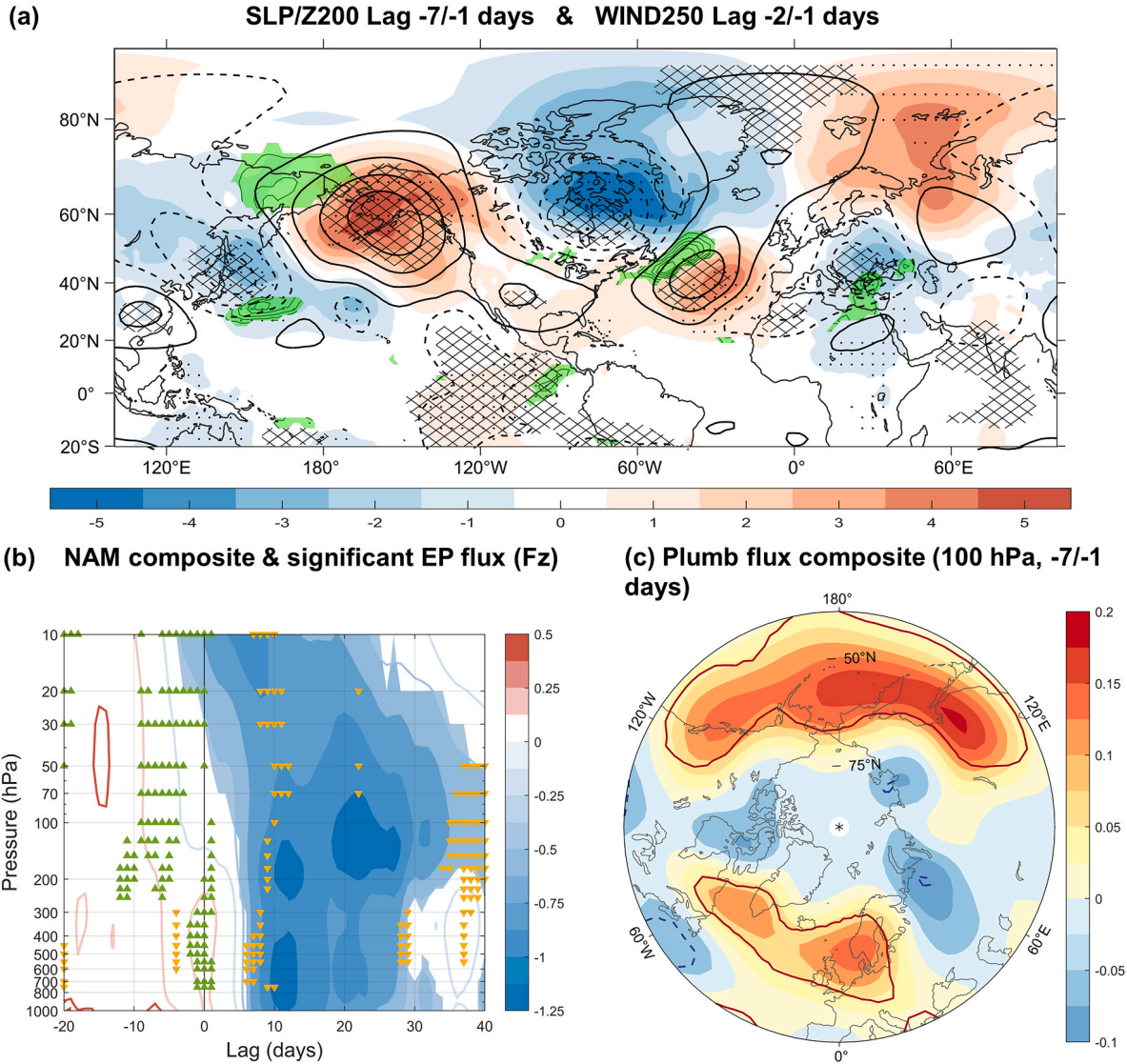

**Fig. 2. P Analogues' composite maps.** (a) Persistent (P) events from daily lags −2 to −1: Wind speed at 250 hPa anomalies (green shaded in m s$^{-1}$, only positive significant anomalies at 95% conf. int. are shown). For daily lags −7 to −1: SLP anomalies (shadings in hPa, 95% confidence interval in stippling) and geopotential height at 200 hPa (black contours in gpm, 95% conf. int. in hatch) anomalies; contours start at 100 gpm and are drawn every 200. (b) Composite daily evolution for P events (lags −20 to +40) of standardized NAM index (contours and 95% conf. int. in shaded) and vertical component of planetary (k = 1–3) Eliassen-Palm flux anomalies north of 50°N (upward/downward EP flux significant at 95% conf. int. shown with upward green/downward orange triangles). (c) Vertical component of Plumb (1985) flux at 100 hPa averaged from days −7 to −1, with 95% conf. int. inside the contours. (For interpretation of the references to color in this figure legend, the reader is referred to the Web version of this article.)

The sequence of dynamical events downstream from the North Pacific, starting with the anomalous high and following with the induced low over North America that together with the North Atlantic anticyclone drive the jet state, also shows remarkable differences. When comparing both Plumb Flux figures (Figs. 2c and 3c), it seems that the main initial driver of the P events is the atmospheric circulation over the North Pacific. The anomalous high is strong enough to activate the global WN2 wave, which places the anomalous low over North America (Fig. 2a). This, combined with the North Atlantic anticyclone, tilts the jet meridionally and allows downstream North Atlantic cyclones to drag much warm air to northern latitudes in P events. This results in qualitatively different RWB, with a more important meridionally-oriented tongue of high potential temperature values reaching further north at 2 PVU (Fig. S4b), potentially contributing to the greater NAO- persistence. For NP, the Pacific North America-like pattern (PNA) places the high over the Western Pacific (Fig. 3a), which is responsible for the significant Plumb flux in Fig. 3c.

However, this particular configuration of the low over North America together with the small and western North Atlantic anticyclone leads to a more zonally oriented jet, less upward propagation in connection with the RWB and no important northward advection of warm air (Figs. S4b and S4c). Therefore, both Pacific and Atlantic processes appear to contribute to NAO- persistence in P events.

To further understand the origin of the SLP/z200 anomalies in the Northern Hemisphere (particularly in the North Pacific) the days before P and NP events (Fig. 2a and 3a), composite global anomalies for SST and z200 are plotted for −23/-8 lagged days (Fig. 4). Due to the criteria applied for event selection (section 3.2), La Niña signal is apparent in the Pacific in both subsets. Both SST patterns present a very similar intensity in the equatorial Pacific. The NP population shows slightly lower mean SST anomaly than the P one (not statistically significant), −1.11 K and −0.91 K, respectively. The Gill response of NP events is westward displaced compared to P events. The latter could be related to the different shape of the SST anomalies, since La Niña of P events is warmer in the west and colder in the east, than La Niña of NP events (not shown). For P events (Fig. 4a), La Niña induces a Gill-type response in the tropical Pacific z200 anomaly field, and an arched wave pattern toward the North Atlantic (Fig. S3c). The wave locates an anticyclone over the North Atlantic weeks before P event dates, which seems to be necessary for the occurrence of the events. La Niña also induces a Gill-type response for NP events (Fig. 4b), but does not appear to propagate into the extratropics. There is no evident and statistically significant wave pattern connecting the Pacific and Atlantic. It is not until daily lags −7/-1 when the atmospheric link between the two basins is established (Fig. 3a).

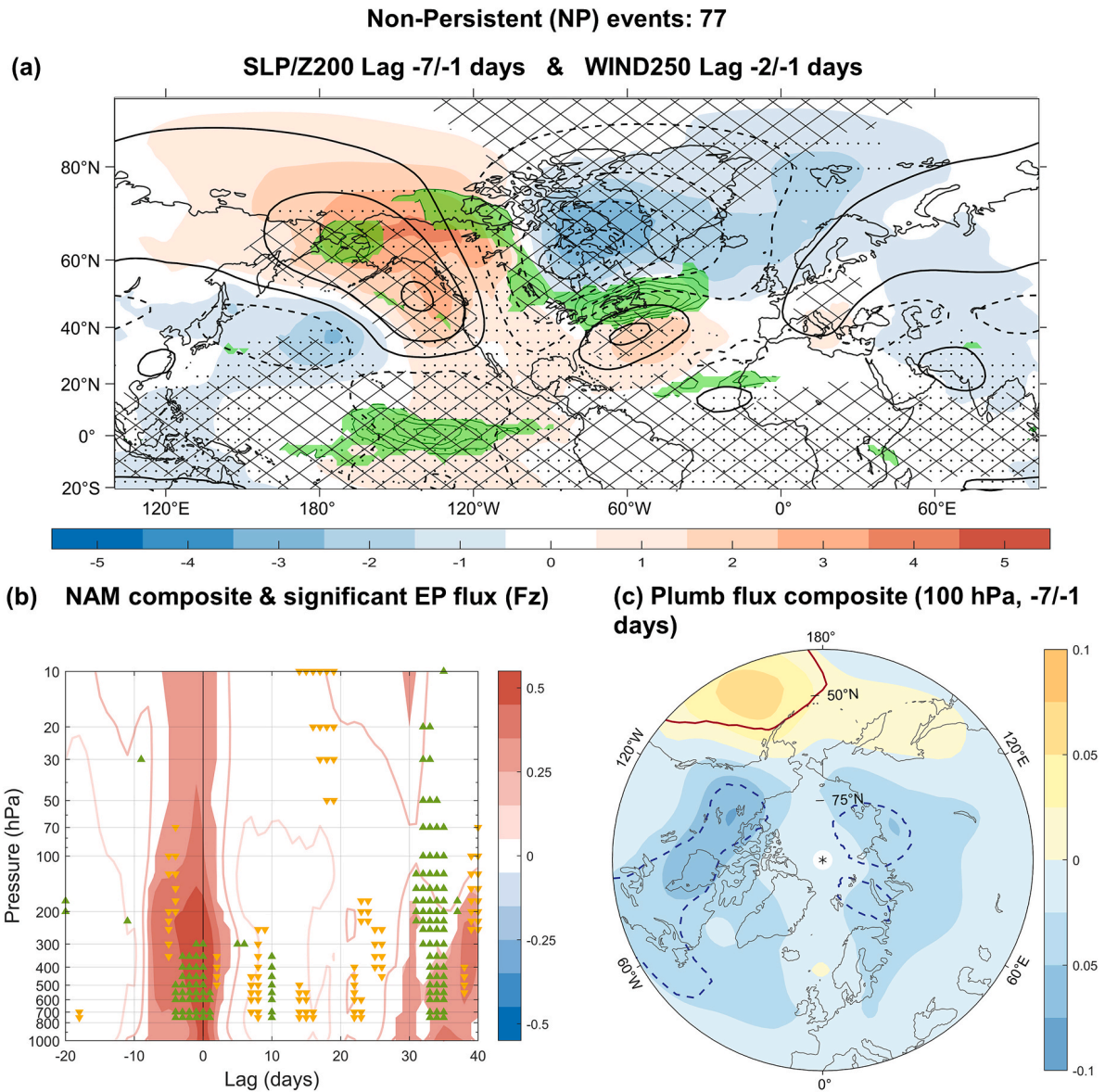

**Fig. 3. NP Analogues' composite maps.** (a) Non-Persistent (NP) events from daily lags −2 to −1: Wind speed at 250 hPa anomalies (green shaded in m s$^{-1}$, only positive significant anomalies at 95% conf. int. are shown). For daily lags −7 to −1: SLP anomalies (shadings in hPa, 95% confidence interval in stippling) and geopotential height at 200 hPa (black contours in gpm, 95% conf. int. in hatch) anomalies; contours start at 100 gpm and are drawn every 200. (b) Composite daily evolution for P events (lags −20 to +40) of standardized NAM index (contours and 95% conf. int. in shaded) and vertical component of planetary (k = 1–3) Eliassen-Palm flux anomalies north of 50°N (upward/downward EP flux significant at 95% conf. int. shown with upward green/downward orange triangles). (c) Vertical component of Plumb (1985) flux at 100 hPa averaged from days −7 to −1, with 95% conf. int. inside the contours. (For interpretation of the references to color in this figure legend, the reader is referred to the Web version of this article.)

The connection between the tropical Pacific and the North Atlantic takes place differently for P and NP populations. For P events, the connection starts in lag −23/-8 with an arched wave from the tropical Pacific to the North Atlantic. Later, a WN5 circumglobal extratropical Rossby wave establishes at 45° N at lag −7/-1 (Fig. 2a and S3a). On the other hand, NP events establish the connection through a different arched pattern in lag −7/-1 directly (Fig. S3b). This dissimilarity in the connection could be related to the mean flow, since changes in the jet intensity and its meanders affect the direction of Rossby wave propagation. Indeed, the difference between P and NP 250 hPa wind speed composites from lag −30 to +30 days, shows how, between 25° and 45°

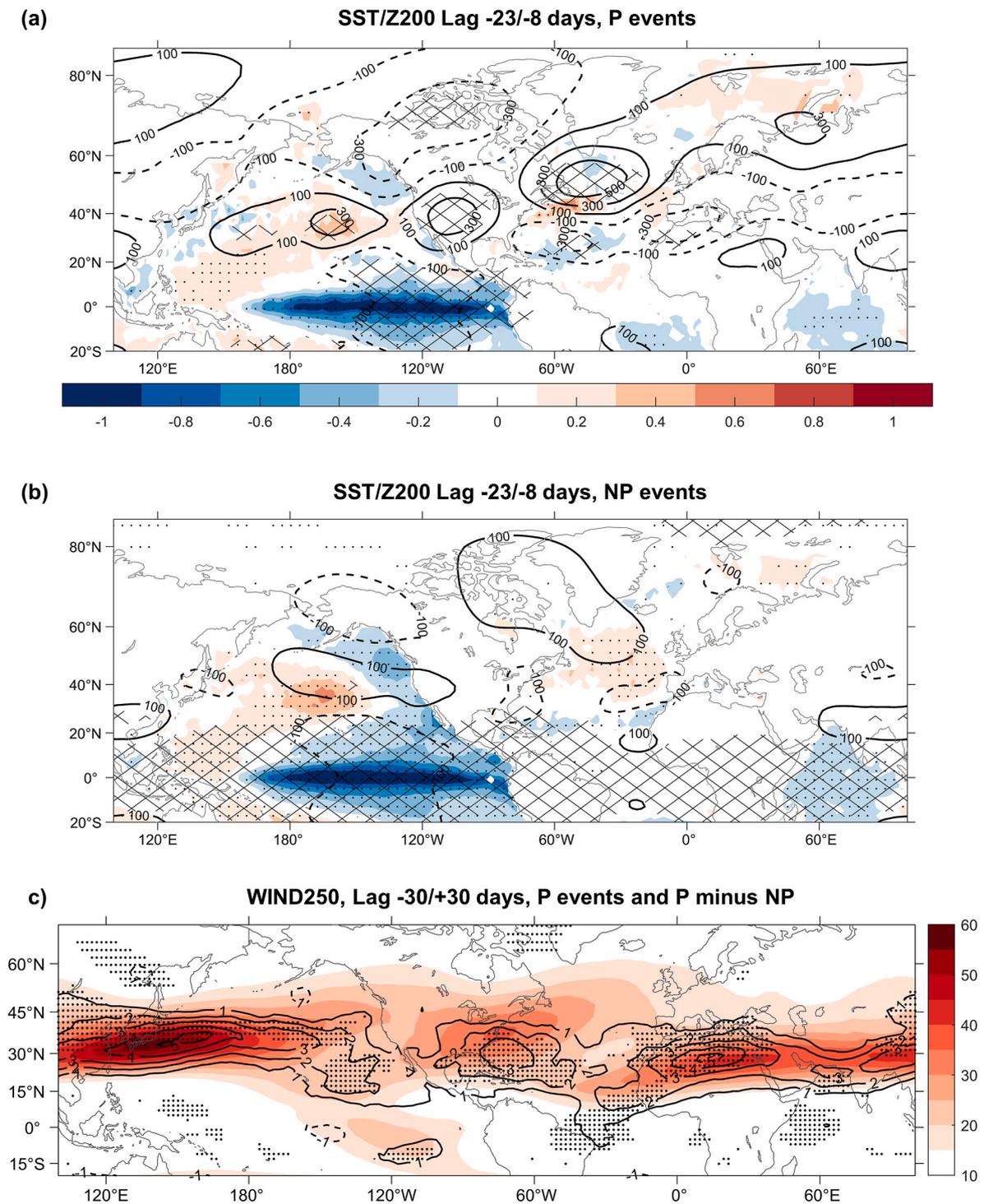

**Fig. 4. P and NP Prior Conditions.** a) Persistent (P) events from daily lag −23 to −8: SST anomalies (shading in K, 95% conf. int. in stippling) and geopotential height at 200 hPa anomalies (black contours in gpm, 95% conf. int. in hatch). (b) as (a) for NP events. (c) P events composites of winter 250 hPa wind from daily lag −30 to +30 (shading in m/s), P minus NP composite difference of winter 250 hPa (contours, in m/s).

N, the jet is stronger in P than in NP along all longitudes except around 120°W, where the jets do not show appreciable differences (Fig. 4c). Additionally, the F18 jet presents an even more pronounced difference to the NP subsample (not shown), consistent with the fact that F18 is the most extreme event in the P sample (Fig S2d). For P events at lag −23/-8, the arched wave pattern linking the Pacific and the Atlantic coincides with the region where the jet is not very strong. Fig. S3c shows some wave activity flux along this pathway. At the same lag for NP events, there is no wave pattern nor wave activity flux (Fig. 4b and S3d). As wave energy does not flow away from the tropics, this can be the reason for the lack of significant z200 anomalies in the extratropics (Fig. 4b). At lag −7/-1, the strong jet associated with P events over the Pacific acts as a waveguide for the wave activity emanating from La Niña. The wave activity emanates from the eastern, middle and western Pacific and travels through the jet towards the North Atlantic where it spreads (Fig. S3a). As a consequence, a circumglobal wave is established some days before P occurrence since the strong jet along all longitudes at 45° N can trap the wave in a circumglobal pattern (Fig. 2a). Conversely, for NP in the absence of a strong jet that trapping the waves, the wave activity only flows from the North Pacific to the North Atlantic through a PNA-like pattern.

## 4. Conclusions

In this study the exceptional weather and climate conditions of late February and March 2018 in the North Atlantic were revisited (Ayarzagüena et al., 2018; Kautz et al., 2019; Drouard et al., 2019; González-Alemán et al., 2022). This period was characterized by an abrupt transition from the positive to the negative phase of the NAO and a subsequent persistence of the negative NAO for several weeks. Evidence is given that tropospheric processes played a decisive role in the occurrence of this North Atlantic event and that of its past analogues in ERA5.

Intraseasonal variation plays a role in allowing abrupt NAO changes in the late winter. Before abrupt NAO transitions, La Niña signal propagates into the extratropics establishing an upper-level anticyclonic circulation in the North Atlantic. The resulting acceleration of the jet stream increased cyclonic activity and RWB, leading to an increased chance of abrupt transitions.

However, only those analogues followed by persistent NAO- presented a circumglobal WN2 a few days before the event. This wave favored the enhancement of upward wave propagation (WN1-3) into the stratosphere from North Pacific and North Atlantic sources. The resulting disturbed stratospheric conditions could also have contributed to the NAO- persistence due to the downward propagation of weak-polar vortex associated anomalies. In addition, poleward cyclonic advection of warm air into the North Atlantic via the large-scale RWB could also have contributed to the persistence. Non-persistent events do not show the same precursors.

Another interesting aspect that deserves further attention is the inhomogeneous distribution in time of the events identified in this study (Table S1). Their frequency dramatically increased after 1995, especially for the P cases. This change coincides with the negative to positive phase transition of the Atlantic Multidecadal Variability (AMV; Knight et al., 2006), which is a regulator of the large-scale circulation in the North Atlantic (Woollings et al., 2012; Gastineau and Frankignoul, 2015; Gómara et al., 2016; Peings and Magnusdottir 2016; Elsbury et al., 2019), the precipitation, and its related extremes, among other aspects (Sutton and Dong, 2012; Simpson et al., 2019). A positive phase of the AMV is known to induce a negative phase of the NAO (Gastineau and Frankignoul, 2015) and it could explain the identified persistent NAO-periods of our study after 1995. Further, the negative NAO phase agrees with an enhancement of extratropical cyclone activity (Gómara et al., 2016; Ruggieri et al., 2021) and precipitation over the Iberian Peninsula during the positive AMV (Simpson et al., 2019), which is consistent with the increase in the frequency of occurrence of the studied events. Another factor that deserves further attention is the possible influence of anthropogenic climate change on this behavior. The results presented here may help to improve extended-range prediction of abruptly developing persistent NAO- states and associated impacts in the North Atlantic (e.g., cold/warm spells, rainfall and wind regimes). To further evaluate contributions of oceanic, tropospheric and stratospheric processes, an analysis of longer-term reanalyses (ERA-20C), model simulations (CMIP6) and seasonal/subseasonal forecasts (Knight et al., 2021; González-Alemán et al., 2022) would be an interesting path for future research.

## Author statement

Marina García-Burgos: Software, Writing-Original draft prparation, Writing- Reviewing and Editing, Formal analysis. Iñigo Gómara: Conceptualization, Methodology, Writing- Reviewing and Editing. Belén Rodríguez-Fonseca: Conceptualization, Methodology, Writing- Reviewing and Editing, Funding acquisition. Pablo Zurita-Gotor: Conceptualization, Methodology, Writing- Reviewing and Editing, Funding acquisition. Juan Jesús Gonzalez-Aleman: Conceptualization, Methodology, Writing- Reviewing and Editing. Blanca Ayarzagüena: Conceptualization, Methodology, Writing- Reviewing and Editing, Funding acquisition.

## Declaration of competing interest

The authors declare that they have no known competing financial interests or personal relationships that could have appeared to influence the work reported in this paper.

## Data availability

Data will be made available on request.

## Acknowledgments, Samples, and Data

Met-Office Hadley Centre HadISST data (https://www.metoffice.gov.uk/hadobs/hadisst/). ECMWF ERA5 data (https://www.ecmwf.int/en/forecasts/dataset/ecmwf-reanalysis-v5). Major SSW dates (https://csl.noaa.gov/groups/csl8/sswcompendium/majorevents.html). This research was funded by the projects EU-H2020 TRIATLAS (no. 817578), Universidad Complutense de Madrid FEI-EU-19-09, Spanish-MINECO PRE4CAST (CGL2017-86415-R) and JeDiS (RTI2018-096402-B-I00). Also supported by the Grant PRE2019-090618 from Spanish Ministerio de Ciencia e Innovación. We thank Joaquim G. Pinto (KIT, Germany) for providing the cyclone tracking algorithm. We also thank Bernat Jiménez-Esteve and another two anonymous reviewers for their pertinent comments and suggestions, which have contributed to improve this manuscript.

## Appendix A. Supplementary data

Supplementary data to this article can be found online at https://doi.org/10.1016/j.wace.2023.100609.